\documentclass[a4paper,fleqn,usenatbib,useAMS,usefloat,referee ]{mnras}
\usepackage{mathptmx}
\usepackage[normalem]{ulem}
\usepackage[T1]{fontenc}
\usepackage{ae,aecompl}
\usepackage{longtable}

\usepackage{graphicx}	
\usepackage{epstopdf}
\usepackage{xcolor}
\usepackage{amsmath}	
\usepackage{amssymb}	
\usepackage{subfig}
\usepackage{ulem}
\usepackage{booktabs}
\usepackage{float}
\usepackage{array}
\usepackage{caption}
\usepackage{scalerel}
\title[HR 6819 might not contain a dormant black hole]
{Does the HR 6819 triple system contain a dormant black hole? Not necessarily}
\author[ Mazeh and Faigler]{
\newauthor{ Tsevi Mazeh and Simchon Faigler} 
\\
{School of Physics and Astronomy, Faculty of Exact Sciences,}\\
{Tel Aviv University, Tel Aviv  69978, Israel}
}
\date{Accepted XXX. Received YYY; in original form ZZZ}

\pubyear{2019}

\begin{document}
\label{firstpage}
\pagerange{\pageref{firstpage}--\pageref{lastpage}}
\maketitle

\begin{abstract}
A recent paper by Rivinius et al.~proposed that HR 6819 is a triple system, with a distant Be star and a binary of 40-day orbit, composed of a B3 III giant and a dormant black hole (BH). We suggest that the evidence for this model is not conclusive. 
In an alternative model, the companion of the giant is by itself a short-period binary in, say, a $\sim$$4$-day orbit, consisting, for example, of two A0 stars. Each of the two A0 stars should contribute $\sim$$4\%$ of the total brightness of the system in the $V$ band, and their spectral lines are moving due to their assumed orbital motion with an unknown period.  Therefore, only a careful analysis of the observed spectra can exclude such a model. Before such an analysis is presented and upper limits for the depths of the hypothetical A0 star absorption lines are derived, the model of a hidden close pair is more probable than the BH model.
\end{abstract}

%
\section{Introduction}

HR 6819 is a nearby early-type Be star at a distance of $\sim$$350$ pc. As reviewed  by \citet[][Ri20]{rivinius20} in a recent intriguing paper, \citet{maintz03}  observed a series of sharp lines in the spectrum of HR 6819 that do not come from the Be star, but instead are emanating from another star, B3 III, in the system. The sharp lines revealed an orbital period of 40 days, turning the system into a hierarchical triple system,  [(?+B3 III)+Be], with the Be star as the distant companion.
Additional 51 spectra obtained by Ri20 and careful analysis revealed an amplitude of $\sim$$60$ km/s of the B3 III star.

From the RV amplitude, together with some estimate of the giant mass, which Ri20  claim to be at least $\sim$$5\,M_{\odot}$, they conclude that the mass of the unseen companion is at least $4.2\, M_{\odot}$. 
The key element of the argument of the Ri20 paper is that the unseen companion cannot be a main-sequence (MS) star. 
A $4.2\,M_{\odot}$ MS B star should have $\sim$$10\%$ of the system luminosity and its Balmer lines can have a depth of $\sim$$60\%$ relative to its own continuum. Consequently, those lines should have a  depth of $\sim$$6\%$ relative to the combined continuum of the system. We note in passing that the location of the lines of the presumed $4.2\,M_{\odot}$ star is well known as a function of the 40-day orbital phase, making their detection easier. Nevertheless, such lines have not been seen in the many spectra obtained by the authors. Therefore, the Ri20 argument goes, the unseen object must be a BH. As no X rays were observed from the system, the BH is dormant, with no supply of material from its giant companion.

BHs, dormant BHs in particular, are quite rare. See, for example, a report about the large search of the LAMOST spectroscopic database by \citet{zheng19}, and a recent discussion of the Ri20 suggestion by \citet{loeb20}, published after this paper was submitted. Consequently, one needs to check carefully if any other model for HR 6819, which might be more probable, is consistent with the observations. This short communication suggests that such a model exists, and some additional observations and/or analysis should be done before one can conclude that the system contains a BH.

%
\section{Alternative model}

\begin{figure*} 
	\centering
	\resizebox{15cm}{7cm}
	{		\includegraphics{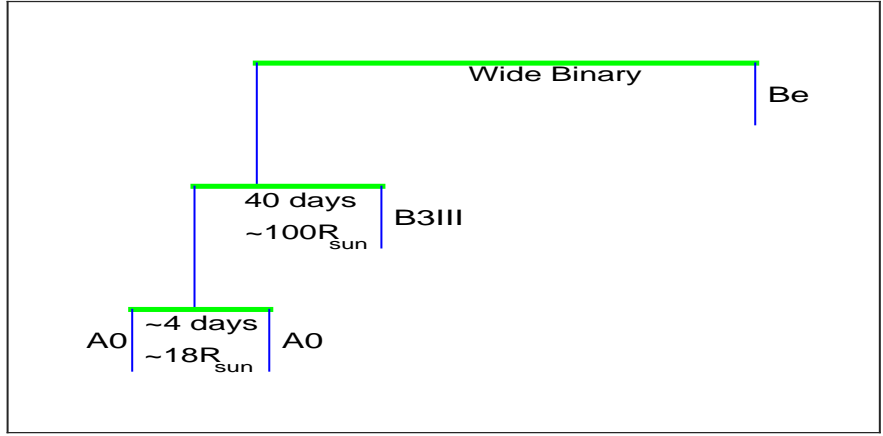}  	}
	\caption{Mobile diagram of the proposed HR 6819 quadruple system. Diagram not to scale.}
	\label{fig:mobile}
\end{figure*}

We propose an alternative model for the HR 6819 system---the unseen companion is by itself a close binary of, say, a $4$-day orbital period. This makes the HR 6819 a hierarchical quadruple, composed of a distant Be star and a hierarchical triple, with the B3 III in a 40-day orbit around the close pair. 
The close binary could be composed, for example, of two A0 stars of 
$\sim$$2.3\,M_{\odot}$ \citep[e.g.,][]{pecaut13}, consistent with the RV modulation of the B3 III; see Figure~\ref{fig:mobile} for a mobile diagram of the suggested system.

%

We know of quite a few similar quadruple systems \citep[see, for example, a discussion by][]{hamers20}. 
The period ratio of  10:1 we suggest for the close triple, [(A0+A0)+B3 III], assures that it is dynamically stable, especially because the outer orbit, that of the B3 III star, is very close to being circular. \citet{tokovinin08} considered  a minimum ratio of 5:1 for a triple to be dynamically stable, and his catalog contains quite a few triples close to this ratio. In fact, the very existence of the giant as a distant companion to the two A-star system could push the pair into a close orbit by eccentricity pumping and relative inclination modulation  \citep[e.g.,][]{mazeh79} through the Kozai-Lidov effect, as suggested by \citet{fabrycky07}.

To estimate the relative luminosity of the A0 stars we use the apparent brightness of the system,  $V=5.36$, and its distance, $\sim$$350$ pc (see Ri20 discussion), to obtain $M_V\sim$$-2.36$. 
The absolute magnitude of each of the two hypothetical A0 stars should be
 $M_V\sim$$1.1$ \citep[e.g.,][]{pecaut13}. Ignoring interstellar extinction,
this 
%
results in a contribution of $\sim$$4\%$ of each of the A0 stars 
 to the luminosity of the HR 6819 system in the $V$ band. 
A similar calculation for the $4.2M_{\odot}$ presumed star considered by Ri20 results in a relative contribution of $\sim$$18\%$ and a 
Balmer-line depth of $\sim$$10\%$.

We note that 
\begin{itemize}
\item
A stars typically have absorption lines that can reach hundreds of km/s broadening \citep[e.g.,][]{zorec12}, 
\item
their typical Balmer line depth can attain a depth of $\sim$$50\%$, and
\item
in our model, the location of the lines changes due to the RV modulation of the A0 stars in their orbital motion with an unknown orbital period. The amplitude of their RV modulation in our specific model is $\sim$$100$ km/s.
\end{itemize}

These three features make the detection of the absorption lines of the presumed A0 stars quite challenging, given their $\sim$$2\%$ depth relative to the combined continuum of the system, and the presence of the absorption lines of the two B stars and the emission lines of the Be star. 

There are two key differences between the two A0 model and the $4.2M_{\odot}$-star model rejected by Ri20. First, the Balmer lines of the A0 stars are of $\sim$$2\%$ depth, whereas the lines of the $4.2M_{\odot}$ star should have $\sim$$10\%$ depth. Second, the lines of the two A0 stars are moving with a high amplitude and  an unknown period, while the position of the  lines of the rejected model are well known.

Therefore, although the $4.2M_{\odot}$-star model is rejected by Ri20, the two A0 model is still viable.  Before a search in the observed spectra for the two A0 absorption lines is carefully done and accurate upper limits are derived, the burden of proving that a BH is hidden in the HR 6819 system is not lifted.

A close pair like we suggest has a separation of 
$\sim$$18\,R_{\odot}$. With the two stars of $\sim$$2\,R_{\odot}$, the system should  display eclipses, unless its orbital inclination is smaller than $\sim$$75^{\circ}$. A full eclipse should have a depth of $\sim$$3\%$. The TESS light curve of HR 6819 (TIC 118842700) could  easily show such an eclipse, despite the erratic modulation of the system, as shown by Ri20 who searched for a periodic sinusoidal signal. In any case, the fact that an eclipse is not seen suggests that the inclination of the presumed binary is less than  $\sim$$80^{\circ}$; below that angle any eclipse should be quite shallow.

The same consideration goes for a possible eclipse of the B3 III by the A stars. 
The separation of the 40-day orbit is somewhat larger than 100 $R_{\odot}$. An eclipse is unlikely, unless the inclination is larger than $\sim$$80^{\circ}$. 
On the other hand, in our model the 40-day orbit cannot have a too low inclination, below, say, $60^{\circ}$, otherwise the masses of the close pair get too large, and their luminosity too bright for hiding their absorption lines in the spectra.
We therefore conclude that even if we assume coplanarity for the close-pair and the 40-day orbits, our model allows for inclination in the range between $65^{\circ}$ and $80^{\circ}$. 
The probability for such an inclination range is $\sim$$0.25$, assuming an isotropic distribution of the  angular-momentum vector of the system.
Assuming relative inclination between the two orbital planes, which is not so surprising \citep[e.g.,][]{tokovinin08}, makes our model less restrictive.

%
\section{Conclusion}
We have shown that a viable model for HR 6819, consistent with the available observations, is a quadruple system, with a close pair composed of two A0 stars, instead of the BH conjecture. 
%
We hope that  the coming paper of Hadrava et al., reported by Ri20, which is going to include a careful analysis of the spectra, can shed some light on the nature of the companion of the B3 III star. 

Furthermore, the separation between the two components of the 40-day binary is $\sim$$1.5$ milli-arcsec (mas). At such separation, Gravity \citep{gravity17} and maybe Gaia \citep[e.g.,][]{gaia16} could resolve the A0-star binary from their B3 III giant  companion, as their combined luminosity is $\sim$$10\%$ of that of the giant. Alternatively, a few precise astrometric measurements might be able to detect the orbital motion of the giant, as its amplitude is expected to be of $\sim$$0.75$ mas, and its orbital period and eccentricity are well known.

While writing this short letter we noticed a recent publication by \citet[][van20]{vandenheuvel20} with regard to the suggestion that the red giant 2MASS J05215658+4359220 has a BH companion \citep{thompson19}. Similarly, van20 conjecture that the unseen companion in that system is a close binary. 
Questions have also been raised \citep{el-badry20,abdul20,eldridge19} about the identification of a $70\,M_{\odot}$  BH as a companion to the B star by \citet{liu19}, but see \citet{liu20} for a rebuttal.
It seems that reaching the goal of confidentially finding dormant BHs in close binaries, as companions to giants in particular, is still further ahead.

\section*{Data Availability}
%
No new data were generated or analysed in support of this research.

\section*{Acknowledgments}
%
 We are indebted to the referee for extremely useful suggestions. We thank Micha Engel for his kind assistance, and Laurent Eyer and Berry Holl for carefully reading the paper and for their comments.
This research was supported by grant No.~2016069 of the United States-Israel Binational Science Foundation (BSF) and by the 
grant no.~I-1498-303.7/2019 of the German-Israeli Foundation.


\bibliographystyle{mnras}
\bibliography{HR_6819}

\end{document}